\documentclass{jpsj2}

\def\runtitle{The real-space renormalization group applied to 
diffusion in inhomogeneous media}
\def\runauthor{Mitsuhiro \textsc{Kawasaki}}

\title{%
The Real-Space Renormalization Group \\ Applied to 
Diffusion in Inhomogeneous Media
}

\author{%
Mitsuhiro \textsc{Kawasaki}\thanks{mituhiro@riam.kyushu-u.ac.jp}
}

\inst{%
Research Institute for Applied Mechanics, Kyushu University, Kasuga 816-8580
}

\recdate{\today}

\abst{%
The real-space renormalization group technique is introduced to evaluate the 
effective diffusion constant for diffusion in inhomogeneous media, 
which has been obtained by singular perturbation methods. 
Our method is formulated on a discretized real space and hence 
it can be easily combined with numerical studies for partial 
differential equations.
}

\kword{%
real-space renormalization group, effective diffusion constant, 
inhomogeneous media
}

\begin{document}
\sloppy
\maketitle

\section{Introduction}
\label{sec-intro}

The renormalization group (RG) is one of general techniques to study 
macroscopic (or self-similar) and universal characteristics of 
equilibrium and non-equilibrium systems whose microscopic models 
have been known \cite{Goldenfeld}. 

In the present paper, the RG technique is applied to the problem of diffusion 
in inhomogeneous media. It can be considered as one of pedagogical examples of 
the RG method. Furthermore, since our real-space RG method is formulated on a 
discrete space, our solution can provide a simple example to show that the 
RG method and numerical simulations of partial differential equations 
can be easily combined. 

For example, there are some attempts to apply the RG technique to Navier-Stokes
equation \cite{Forster}. It is well known that the scale of turbulent flows 
of any practical significance are expected to be much larger than the 
Kolmogorov scale, which is the smallest scale of activated eddies.
Hence, in order to save computer resource, only some large-scale eddies are 
computed explicitly and effects from other smaller eddies are modeled as a 
eddy viscosity \cite{Kaneda}. 
The numerical simulations based on the idea of sub-grid 
modeling are called the large eddy simulations. 
The eddy viscosity used in large eddy simulations 
for turbulent fluids is expected to be evaluated with the RG 
applied to Navier-Stokes equation \cite{McComb}.

The only difference between the applications of 
the RG to Navier-Stokes equation and the present application to diffusion 
is whether the equations to be solved for decimation are non-linear or linear. 
So, our solution can be expected to help investigate applicability of the RG 
technique to solve other partial differential equations with aid of 
numerical computation.

This introductory section is concluded with note that the problem has been 
solved with other singular perturbation methods like multi-scale methods 
\cite{Bensoussan}. 

The present paper is organized as follows: 
After the problem to be studied is explained in \S\ \ref{sec-problem}, 
the RG is formulated in \S\ \ref{sec-rg} step by step. 
At first, the diffusion equation is extended formally to close 
the renormalization transform in \S\ \ref{sec-extension}. 
The construction of the renormalization 
transform is divided into two procedures, 
decimation explained in \S\ \ref{sec-decimation} and rescaling in 
\S\ \ref{sec-rescaling}.
The formulae of renormalization transform are obtained in \S\ \ref{sec-rt}. 
With the formulae, the fixed points are evaluated in 
\S\ \ref{sec-fixed-points}. Finally, the effective diffusion equation 
is obtained in \S\ \ref{sec-eff}. 
The results are summarized and discussed in \S\ \ref{sec-summary}.

\section{Formulation of the Problem}
\label{sec-problem}

A one-dimensional inhomogeneous medium where the diffusion constant $D(x)$ 
is a spatially periodic function is assumed. 
The period is denoted as $l$ and hence $D(x+l) = D(x)$. 
The objective of our study is evaluation of the effective diffusion constant 
$D_e$ when the spatial resolution of observation of diffusion process is 
much larger than the period $l$. The scale of observation is denoted as 
$L_0$. The diffusion equation is written as 
\begin{equation}
\frac{\partial}{\partial t}\rho(x,t) = \frac{\partial}{\partial x}D(x)
\frac{\partial}{\partial x}\rho(x,t), \ \ \ \ D(x+l) = D(x).
\label{diffusion-equation}
\end{equation}

In order to make formulation of the real-space RG easier, 
we discretize the continuum space into a chain \cite{footnote-discretization}. 
The final results for the effective diffusion constant are obtained in 
the continuum limit in \S\ \ref{sec-eff}.

Grid points on the chain are numbered with the integers $\{\cdots,i-1,i,i+1,
\cdots \}$ and the lattice constant is assumed to be $a \ (\ll l)$, i.e., 
$x = \lim_{a \rightarrow 0} i a$.
There are $N_l$ sites in one period $l$, i.e., $N_l \equiv l/a$. 
The probability $P_i(t)$ that a random walker is found at site $i$, or 
the number of the random walkers, is related to 
the probability density $\rho(x,t)$ as 
\begin{equation}
\rho(x,t) \equiv \lim_{a \rightarrow 0} \frac{P_i(t)}{a}.
\label{definition-discrete-probability}
\end{equation}
Then, the diffusion equation (eq.\ (\ref{diffusion-equation})) is discretized 
as 
\begin{equation}
\frac{d P_i(t)}{dt} = \frac{1}{a^2} \left[ 
D_{i-1}P_{i-1}(t)-(D_{i-1}+D_i)P_i(t)+D_i P_{i+1}(t) \right].
\label{discrete-diffusion-equation}
\end{equation}

\section{Renormalization Group}
\label{sec-rg}

In this section, we formulate the renormalization transform for 
the discretized diffusion equation (eq.\ (\ref{discrete-diffusion-equation})) 
and obtain the fixed point. 
The construction of the renormalization transform is explained by 
being divided into two steps, decimation and rescaling.

\subsection{Formal extension of the diffusion equation}
\label{sec-extension}

Before construction of the renormalization transform, 
new parameters are introduced in 
the discretized diffusion equation (the master equation) in order 
to obtain the closed renormalization transform as follows:
\begin{eqnarray}
\frac{d P_i(t)}{dt} & = & \frac{1}{a^2} \int_0^t d\tau 
\left[ D_{i-1}(t-\tau)P_{i-1}(\tau)-(D_{i-1}(t-\tau)+D_i(t-\tau) 
\right. \nonumber \\
&& + \left.
U_i(t-\tau))P_i(\tau)+D_i(t-\tau) P_{i+1}(\tau) \right].
\label{extended-descrete-diffusion-equation}
\end{eqnarray}
Although the memory effect is introduced, the effect does not exist in fact. 
Hence, the memory function $D_i(t)$ is defined with the delta function as 
\begin{equation}
D_i(t) = \delta(t) D_i \ \ \ \ \mbox{where} \ 
\int_0^t d\tau \delta(t-\tau) P_i(\tau) \equiv P_i(t).
\end{equation}
Furthermore, since $U_i(t)$ is introduced only for closing the 
renormalization transform, $U_i(t)=0$.

In order to make decimation procedure easier, the integro-differential 
equation, eq.\ (\ref{extended-descrete-diffusion-equation}), is 
Laplace transformed as 
\begin{eqnarray}
s P_i(s)-P_i(0) & = & \frac{1}{a^2} \left[
D_{i-1}(s)P_{i-1}(s)-(D_{i-1}(s)+D_i(s)+U_i(s))P_i(s) \right. \nonumber \\
&& \left. +D_i(s) P_{i+1}(s) \right].
\label{laplace-transform-master-eq}
\end{eqnarray}
Here, we denote the Laplace transform of $P_i(t), D_i(t), U_i(t)$ as 
$P_i(s), D_i(s), U_i(s)$ respectively. The Laplace variable $s$ is 
explicitly written for prevention of confusing.
For simpler notation, eq.\ (\ref{laplace-transform-master-eq}) is 
rewritten as 
\begin{eqnarray}
s P_i(s)-P_i(0) & = & \left[
w_{i-1}(s)P_{i-1}(s)-(w_{i-1}(s)+w_i(s)+v_i(s))P_i(s) \right. \nonumber \\
&& \left. +w_i(s) P_{i+1}(s) \right],
\label{laplace-transform-master-eq-2}
\end{eqnarray}
where the Laplace transformed jump rates are defined as 
\begin{equation}
w_i(s) \equiv D_i(s)/a^2, \ \ v_i(s) \equiv U_i(s)/a^2.
\end{equation}

\subsection{Decimation}
\label{sec-decimation}

Decimation procedure, i.e., reduction of the degrees of freedom, 
is performed in this subsection. It is the first step of the renormalization 
transform.

Especially in spin systems on lattices, the procedure is often implemented 
as averaging the variables in small blocks which the whole system is divided 
into \cite{Kadanoff}. On the other hand, for diffusion phenomena, 
development for long time is regarded as a kind of average process for 
distribution of the random walkers. 
Hence, we expect that the explicit average is not needed and 
reduce the degrees of freedom with algebraic elimination explained below. 
It means that the effective development equation for the non-averaged 
probability density $\rho(x,t)$, instead of the averaged density, 
is obtained at last. 

Although it appears that probability is annihilated with decimation, 
the renormalization transform for the initial condition, 
eq.\ (\ref{conservation-probability}), shows that 
the total probability (or the total amount of matter) is conserved 
in the renormalization transform. 

\begin{figure}[hptd]
\begin{center}
\includegraphics[width=10cm,keepaspectratio]{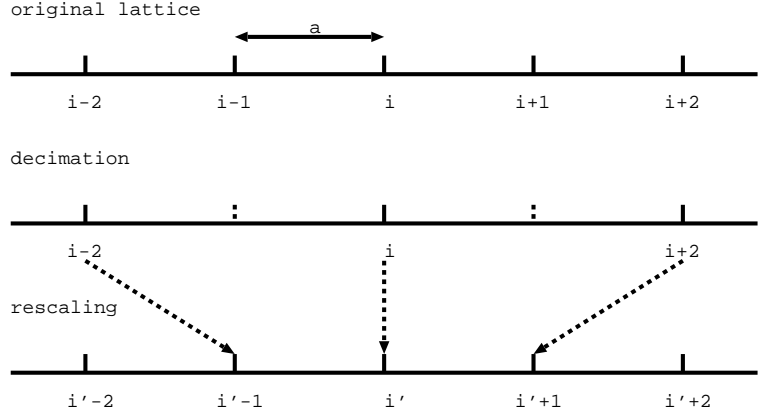}
\caption{Schematic illustration of decimation and rescaling procedure 
performed on the discrete chain.}
\label{figure-schematic-explation}
\end{center}
\end{figure}

Decimation process is performed by just eliminating the probability 
of the nearest neighbor sites, $P_{i-1}(s)$ and $P_{i+1}(s)$, 
from the development equation for the probability $P_i(s)$, 
eq.\ (\ref{laplace-transform-master-eq-2}). 
Decimation process as well as rescaling is explained schematically in 
Fig.\ \ref{figure-schematic-explation}.

Elimination of $P_{i-1}(s)$ is done by solving 
eq.\ (\ref{laplace-transform-master-eq-2}) 
where the site index is changed as $i \rightarrow i-1$. 
$P_{i-1}(s)$ is given as 
\begin{equation}
P_{i-1}(s) = \frac{w_{i-2}(s) P_{i-2}(s)+w_{i-1}(s) P_{i}(s)}
{s+w_{i-2}(s)+w_{i-1}(s)+v_{i-1}(s)}+
\frac{P_{i-1}(0)}{s+w_{i-2}(s)+w_{i-1}(s)+v_{i-1}(s)}.
\label{decimation-eq}
\end{equation}
The probability $P_{i+1}(s)$ is obtained in the same way. 
By inserting $P_{i-1}(s)$ and $P_{i+1}(s)$ in 
eq.\ (\ref{laplace-transform-master-eq-2}), the decimated master equation 
is obtained as 
\begin{eqnarray}
&& s P_i(s)-P_i(0)-\frac{w_{i-1}(s)}{s+w_{i-2}(s)+w_{i-1}(s)
+v_{i-1}(s)}P_{i-1}(0) \nonumber \\
&& -\frac{w_i(s)}{s+w_{i}(s)+w_{i+1}(s)+v_{i+1}(s)}P_{i+1}(0) \nonumber \\ 
& = & \frac{w_{i-2}(s)w_{i-1}(s)}{s+w_{i-2}(s)+w_{i-1}(s)
+v_{i-1}(s)}P_{i-2}(s) \nonumber \\
&& -\left[ \frac{w_{i-2}(s)w_{i-1}(s)}{s+w_{i-2}(s)+w_{i-1}(s)+v_{i-1}(s)} 
\right. \nonumber \\
&& \left. +\frac{w_{i}(s)w_{i+1}(s)}{s+w_{i}(s)+w_{i+1}(s)+v_{i+1}(s)} 
\right. \nonumber \\
&& \left. +v_i(s)+\frac{s w_{i-1}(s)+w_{i-1}(s) v_{i-1}(s)}{s+w_{i-2}(s)
+w_{i-1}(s)+v_{i-1}(s)} \right. \nonumber \\
&& \left. +\frac{s w_{i}(s)+w_{i}(s) v_{i+1}(s)}{s+w_{i}(s)+w_{i+1}(s)
+v_{i+1}(s)} \right] P_i(s) \nonumber \\
&& +\frac{w_{i}(s)w_{i+1}(s)}{s+w_{i}(s)+w_{i+1}(s)+v_{i+1}(s)}P_{i+2}(s)
\label{coarse-grained-master-equation}
\end{eqnarray}

Only the probabilities of the nearest neighbor sites appear in the 
master equation (eq.\ (\ref{laplace-transform-master-eq-2})). 
This is the reason why decimation is performed easily and it is 
locality of the renormalization transform.

\subsection{Rescaling}
\label{sec-rescaling}

Next, we perform rescaling procedure, which is the second step of the 
renormalization transform. 
Since one of every two neighboring sites is eliminated, 
the sparse grid points are renamed as 
\begin{equation}
i \rightarrow i' \equiv i/2,
\label{lattice-index-rescaling}
\end{equation}
where $i$ is assumed even. In the continuum limit, the rescaling 
eq.\ (\ref{lattice-index-rescaling}) corresponds to 
$x \rightarrow x' \equiv x/2$, since the lattice constant $a$ is not rescaled.

Since the spatial scale is rescaled, time development slows down accordingly. 
It is expressed as 
\begin{equation}
t \rightarrow t' \equiv 2^{-\mu} t,
\label{time-rescaling}
\end{equation}
where the exponent $\mu$ is a positive number to be determined later. 
It means that the Laplace variable is rescaled as 
\begin{equation}
s \rightarrow s' \equiv 2^{\mu} s.
\label{laplace-variable-rescaling}
\end{equation}

Properly speaking, the probabilities $\{P_i(t)\}$ are also to be rescaled. 
However, the diffusion equation that we consider is linear and hence 
it does not matter whether the probabilities are rescaled. 
For simplicity, we do not rescale the quantities;
\begin{equation}
P_i(t) \rightarrow {P'}_{i'}(t') \equiv P_i(t).
\label{probability-rescaling}
\end{equation} 
Then, by using rescaling of the Laplace variable 
(eq.\ (\ref{laplace-variable-rescaling})) and the definition of the Laplace 
transform, the Laplace transformed probability $P_i(s)$ is rescaled as 
\begin{equation}
P_i(s) \rightarrow {P'}_{i'}(s') \equiv 2^{-\mu} P_i(s).
\label{laplace-transform-probability-rescaling}
\end{equation}

\subsection{The renormalization transform}
\label{sec-rt}

The renormalization transformed jump rates and the initial condition, 
${w'}_{i'}(s'), {v'}_{i'}(s'), {P'}_{i'}(0),$ are 
derived by using the rescaling formulae obtained above. 
By inserting the rescaling formulae (eqs.\ (\ref{lattice-index-rescaling}, 
\ref{laplace-variable-rescaling}, 
\ref{laplace-transform-probability-rescaling})) 
in the decimated master equation 
(eq.\ (\ref{coarse-grained-master-equation})), the renormalized master 
equation is given as 
\begin{eqnarray}
s' {P'}_{i'}(s')-{P'}_{i'}(0) & = & 2^{\mu} \frac{w_{i-2}(s)w_{i-1}(s)}
{s+w_{i-2}(s)+w_{i-1}(s)+v_{i-1}(s)}{P'}_{i'-1}(s') \nonumber \\
&& -\left[ 2^{\mu} \frac{w_{i-2}(s)w_{i-1}(s)}{s+w_{i-2}(s)+w_{i-1}(s)
+v_{i-1}(s)} \right. \nonumber \\
&& \left. +2^{\mu} \frac{w_{i}(s)w_{i+1}(s)}{s+w_{i}(s)+w_{i+1}(s)
+v_{i+1}(s)} \right. \nonumber \\
&& \left. +2^{\mu} \left( v_i(s)+\frac{s w_{i-1}(s)+w_{i-1}(s) v_{i-1}(s)}
{s+w_{i-2}(s)+w_{i-1}(s)+v_{i-1}(s)} \right. \right. \nonumber \\
&& \left. \left. +\frac{s w_{i}(s)+w_{i}(s) v_{i+1}(s)}{s+w_{i}(s)+w_{i+1}(s)
+v_{i+1}(s)} \right) \right] {P'}_{i'}(s') \nonumber \\
&& +2^{\mu} \frac{w_{i}(s)w_{i+1}(s)}{s+w_{i}(s)+w_{i+1}(s)+v_{i+1}(s)}
{P'}_{i'+1}(s').
\label{renormalized-master-equation}
\end{eqnarray}
Since the jump rates are the coefficients of the probability ${P'}_{i'}(s')$ 
in the master equation, the renormalized jump rates 
${w'}_{i'}(s'), {v'}_{i'}(s')$ are given as 
\begin{eqnarray}
{w'}_{i'}(s') & = & 2^{\mu} \frac{w_{i}(s)w_{i+1}(s)}{s+w_{i}(s)+w_{i+1}(s)
+v_{i+1}(s)}, 
\label{transition-rate-renormalization-transform1}
\\
{v'}_{i'}(s') & = & 2^{\mu} \left( v_i(s)+\frac{s w_{i-1}(s)+w_{i-1}(s) 
v_{i-1}(s)}{s+w_{i-2}(s)+w_{i-1}(s)+v_{i-1}(s)} \right. \nonumber \\
&& \left. +\frac{s w_{i}(s)+w_{i}(s) v_{i+1}(s)}{s+w_{i}(s)+w_{i+1}(s)
+v_{i+1}(s)} \right).
\label{transition-rate-renormalization-transform2}
\end{eqnarray}
These formulae are the renormalization transform for the jump rates. 

In the renormalized master equation, 
eq.\ (\ref{renormalized-master-equation}), ${P'}_{i'}(0)$ is defined as 
\begin{eqnarray}
{P'}_{i'}(0) & \equiv & P_i(0)
+\frac{w_{i-1}(s)}{s+w_{i-2}(s)+w_{i-1}(s)+v_{i-1}(s)}P_{i-1}(0) \nonumber \\
&& +\frac{w_i(s)}{s+w_{i}(s)+w_{i+1}(s)+v_{i+1}(s)}P_{i+1}(0).
\label{initial-condition-renormalization-transform}
\end{eqnarray}
This is the renormalization transform for the initial condition. 
It is weird that the renormalized initial condition, which should be real, 
depends on the Laplace variable $s$. However, it is shown in eq.\ 
(\ref{simplified-initial-condition-renormalization-transform}) 
that the $s$-dependence vanishes in the limit $l/L_0 \rightarrow 0$. 
Furthermore, for complete rigorous deduction, 
a new parameter should be introduced to 
absorb the $s$-dependence. If the parameter was introduced, 
the $s$-dependence of the renormalized initial condition would be eliminated 
and the newly introduced parameter would become zero in the limit.

By remembering that we are interested in the behavior whose spatial 
scale $L_0$ is much larger than $l$, the renormalization transform formulae 
obtained above are simplified. The characteristic time scale which corresponds 
to the spatial scale $L_0$ is given by the transit time that it takes for 
the random walker to pass the region of length $L_0$. The transit time is 
of the order of ${L_0}^2/D_i$. Converted into the scale of the Laplace 
regime, $s \sim D_i/{L_0}^2$. From the consideration, the order of 
$s/w_i(s)$ is estimated as 
\begin{equation}
\frac{s}{w_i(s)} \sim \frac{a^2}{D_i}s \sim 
\frac{a^2 D_i}{D_i {L_0}^2} \sim \left( \frac{a}{L_0} \right)^2.
\label{order-estimation}
\end{equation}
Since $L_0 \gg a$, $s/w_i(s)$ is a small quantity, $s/w_i(s) \ll 1$.

The estimation of the order is not valid when 
the renormalization transform is performed many times. 
Here, we show that $s/w_i(s)$ is small even at the fixed 
point, i.e., $s^*/{w^*}_{i^*}(s^*) \ll 1$. 
Afterwards, the parameter $\mu$ is determined as $\mu = 1$ and hence 
$s'=2 s$. After the renormalization transform is performed 
$\log_2 N_l$ times, the fixed point is reached. 
Hence, the order is estimated as $s^*/{w^*}_{i^*}(s^*) \sim O(a l/L_0^2)$. 
Since $l \ll L_0$, $s^*/{w^*}_{i^*}(s^*)$ is small. 
It means that $s^{(n)}/{w^{(n)}}_{i^{(n)}}(s^{(n)})$, which is the 
quantity renormalized $n$ times, is small for arbitrary $n$, 
since the rescaled Laplace variable grows monotonically. 

The result of the order estimation means that the jump rates $w_i(s), v_i(s)$ 
can be expand in terms of relatively small parameter $s$. 
The jump rates are expanded up to the first order $O(s)$;
\begin{equation}
w_i(s) = w_i + r_i s+O(s^2), \ \ \ \ v_i(s) = v_i s+O(s^2).
\label{transition-rate-expansion}
\end{equation}
The ratio of the neglected terms to the terms considered explicitly is 
of order of $(a/L_0)^2$ and hence 
contribution from these terms can vanish at the continuum limit. 
Furthermore, the neglected terms at the fixed point are relatively of order 
of $(l/L_0)^2$. It means that the second order quantities can be 
eliminated even at the fixed point, since $L_0 \gg l$.
In addition, note that the initial values for $w_i, r_i, v_i$ are given by 
$w_i \equiv D_i/a^2, r_i = v_i = 0$.

By inserting the expansion (eq.\ (\ref{transition-rate-expansion})) 
in the renormalization transform for $w_i(s), v_i(s)$ 
(eqs.\ (\ref{transition-rate-renormalization-transform1},
\ref{transition-rate-renormalization-transform2})), 
the renormalization transform formulae for $w_i, r_i, v_i$ are derived. 
At first, the transform for $w_i(s)$ is expanded in terms of $s$ as 
\begin{eqnarray} 
& & {w'}_{i'}(s') \nonumber \\
& = & 2^{\mu} \frac{{w_i}(s) w_{i+1}(s)}{s+w_i(s)+w_{i+1}(s)+v_{i+1}(s)} 
\nonumber \\
& = & 2^{\mu} \left[ \frac{w_i w_{i+1}}{w_i+w_{i+1}}
+s \frac{r_{i+1} {w_i}^2+r_i {w_{i+1}}^2-w_i w_{i+1} (1+v_{i+1})}
{(w_i+w_{i+1})^2} \right]+O(s^2).
\end{eqnarray}
On the other hand, ${w'}_{i'}(s')$ is expanded with 
eq.\ (\ref{transition-rate-expansion}) in terms of $s'$ as 
\begin{equation}
{w'}_{i'}(s') \equiv {w'}_{i'} + s' {r'}_{i'} 
= {w'}_{i'} + 2^{\mu} s {r'}_{i'}.
\end{equation}
By comparison, the renormalization transform formulae for $w_i, r_i$ are 
given as 
\begin{eqnarray}
{w'}_{i'} & = & 2^{\mu} \frac{w_i w_{i+1}}{w_i+w_{i+1}}, 
\label{w-renormalization-transform} \\
{r'}_{i'} & = & \frac{r_{i+1} {w_i}^2+r_i 
{w_{i+1}}^2-w_i w_{i+1}(1+ v_{i+1})}{(w_i+w_{i+1})^2}.
\label{r-renormalization-transform}
\end{eqnarray}

Next, by inserting the expansion, eq.\ (\ref{transition-rate-expansion}), 
in the transform for $v_i(s)$, 
eq.\ (\ref{transition-rate-renormalization-transform2}), 
the renormalization transform for $v_i$ is given as 
\begin{eqnarray}
&& {v'}_{i'}(s') 2^{-\mu} \nonumber \\
&=& \left[ v_i+\frac{w_{i-1}(1+v_{i-1})}{w_{i-2}+w_{i-1}}+\frac{w_i(1+v_{i+1})}{w_i+w_{i+1}} \right]s +O(s^2) \nonumber \\
&\equiv& {v'}_{i'} s' 2^{-\mu} \nonumber \\
&=& {v'}_{i'} s.
\end{eqnarray}
By comparison, the renormalization transform of $v_i$ is shown to be 
\begin{equation}
{v'}_{i'} = v_i+\frac{w_{i-1}(1+v_{i-1})}{w_{i-2}+w_{i-1}}+
\frac{w_i(1+v_{i+1})}{w_i+w_{i+1}}.
\label{renormalization-transform-v}
\end{equation}

The renormalization transform for the initial condition is also simplified 
in the similar way. With eq.\ (\ref{transition-rate-expansion}), 
the coefficients in the transform (eq.\ 
(\ref{initial-condition-renormalization-transform})) is expanded as 
\begin{eqnarray}
\frac{w_i(s)}{s+w_{i}(s)+w_{i+1}(s)+v_{i+1}(s)} & = & 
\frac{w_i}{w_i+w_{i+1}}+\frac{r_i w_{i+1}-w_i(1+r_{i+1}+v_{i+1})}
{(w_i+w_{i+1})^2} s \nonumber \\ && +O(s^2).
\label{expansion-coefficient}
\end{eqnarray}
We consider the case that the transform has been performed $n$ times. 
Since $s \sim D_i/L_0^2, {v^{(n)}}_{i^{(n)}}\sim 2^n$, 
the second term on the right hand side of 
eq.\ (\ref{expansion-coefficient}) is estimated as 
\begin{eqnarray}
\frac{{r^{(n)}}_{i^{(n)}} {w^{(n)}}_{i^{(n)}+1}-{w^{(n)}}_{i^{(n)}}
(1+{r^{(n)}}_{i^{(n)}+1}+{v^{(n)}}_{i^{(n)}+1})}{({w^{(n)}}_{i^{(n)}}+
{w^{(n)}}_{i^{(n)}+1})^2}s^{(n)} & \sim & 
\frac{{v^{(n)}}_{i^{(n)}}}{{w^{(n)}}_{i^{(n)}}}s^{(n)} \nonumber \\
& \sim & \frac{a^2{v^{(n)}}_{i^{(n)}}}{a^2{w^{(n)}}_{i^{(n)}}}2^n
\frac{D_i}{L_0^2} \nonumber \\
& \sim & \frac{2^n a^2 {v^{(n)}}_{i^{(n)}}}{L_0^2} \nonumber \\
& \sim & \frac{2^{2 n}a^2}{L_0^2} \nonumber \\
& \leq & \left( \frac{l}{L_0} \right)^2
\label{ordering1}
\end{eqnarray}
By neglecting the quantities of $O((a/L_0)^2), O(a l/L_0^2), O((l/L_0)^2)$, 
the approximation 
\begin{equation}
\frac{{w^{(n)}}_{i^{(n)}}(s)}{s^{(n)}+{w^{(n)}}_{i^{(n)}}(s)
+{w^{(n)}}_{{i^{(n)}}+1}(s)+{v^{(n)}}_{i^{(n)}+1}(s)} \simeq 
\frac{{w^{(n)}}_{i^{(n)}}}{{w^{(n)}}_{i^{(n)}}+{w^{(n)}}_{i^{(n)}+1}}
\end{equation}
is justified.
Hence, the renormalization transform for the initial condition 
eq.\ (\ref{initial-condition-renormalization-transform}) is simplified as 
\begin{eqnarray}
{P^{(n+1)}}_{i^{(n+1)}}(0) & \simeq & {P^{(n)}}_{i^{(n)}}(0)
+\frac{{w^{(n)}}_{i^{(n)}-1}}{{w^{(n)}}_{i^{(n)}-2}+{w^{(n)}}_{i^{(n)}-1}}
{P^{(n)}}_{i^{(n)}-1}(0) \nonumber \\
&& +\frac{{w^{(n)}}_{i^{(n)}}}{{w^{(n)}}_{i^{(n)}}+{w^{(n)}}_{i^{(n)}+1}}
{P^{(n)}}_{i^{(n)}+1}(0).
\label{simplified-initial-condition-renormalization-transform}
\end{eqnarray}
Even though some variables for probability $P_i(s)$ in decimation procedure 
are eliminated, it is shown from 
eq.\ (\ref{simplified-initial-condition-renormalization-transform}) that 
the total probability is conserved as 
\begin{equation}
\sum_{i^{(n+1)}} {P^{(n+1)}}_{i^{(n+1)}}(0) = \sum_{i^{(n)}} 
{P^{(n)}}_{i^{(n)}}(0).
\label{conservation-probability}
\end{equation}

Here, we consider the renormalized initial condition. 
From eq.\ (\ref{simplified-initial-condition-renormalization-transform}), 
the renormalization for the initial distribution localized at the site $m N_l$ 
where $m$ is arbitrary integer is given as 
\begin{equation}
{P^{(n)}}_{i^{(n)}}(0) = \delta_{i^{(n)}, m N_l/2^n}.
\label{renormalized-initial-condition}
\end{equation}
The result is equivalent to the limiting form at the limit 
$t \rightarrow 0$ of the rescaling of $P_i(t)$ 
(eq.\ (\ref{probability-rescaling})).

However, the result of the transform for general initial distribution is 
inconsistent with the limit $t\rightarrow 0$ of rescaling of the 
probability. In order to solve the inconsistency, we would had to reformulate 
decimation procedure with averaging of the neighboring probabilities. 
In the rest of the present paper, we restrict our interest in the localized 
initial distribution like eq.\ (\ref{renormalized-initial-condition}) for 
its simplicity. In the case, there is not any inconsistency.

Since all the formulae are obtained, the value of $\mu$ is determined. 
We consider condition for existence of the non-trivial fixed point 
$w^*$ for the jump rate $w_i$.
From eq.\ (\ref{w-renormalization-transform}), 
assumption that the fixed point $w^*$ exists is written as 
\begin{equation}
w^* = 2^{\mu-1} w^*.
\end{equation}
For $w^*$ is non-zero, it is necessary and sufficient that $\mu = 1$.

Since the number of parameters is increased, we summarize definition and 
the renormalization transform formulae:
\begin{itemize}
\item The master equation
\begin{eqnarray}
s P_i(s)-P_i(0) & = & \left[
w_{i-1}(s)P_{i-1}(s)-(w_{i-1}(s)+w_i(s)+v_i(s))P_i(s) \right. \nonumber \\
&& \left. +w_i(s) P_{i+1}(s) \right],
\label{laplace-transform-master-eq-matome}
\end{eqnarray}
where $w_i(s) \equiv w_i+r_i s, v_i(s) \equiv v_i s, D_i \equiv a^2 w_i$.
\item The renormalization transform 
\begin{equation}
s' = 2 s \ \ \ \ (t' = t/2)
\label{laplace-variable-rescaling-matome}
\end{equation}
\begin{equation}
i' = i/2 \ \ \ \ (x'=x/2)
\label{space-index-rescaling-matome}
\end{equation}
\begin{equation}
{P'}_{i'}(s') = P_i(s)/2 \ \ \ \ 
({P'}_{i'}(t')=P_i(t))
\label{laplace-transform-probability-rescaling-matome}
\end{equation}
\begin{equation}
{w'}_{i'} = 2 \frac{w_i w_{i+1}}{w_i+w_{i+1}}
\label{transition-rate-renormalization-transform-matome}
\end{equation}
\begin{equation}
{r'}_{i'} = \frac{r_{i+1} {w_i}^2+r_i {w_{i+1}}^2-w_i w_{i+1}(1+ v_{i+1})}
{(w_i+w_{i+1})^2}
\label{r-renormaliztion-transform-matome}
\end{equation}
\begin{equation}
{v'}_{i'} = v_i+\frac{w_{i-1}(v_{i-1}+1)}{w_{i-2}+w_{i-1}}+
\frac{w_i(v_{i+1}+1)}{w_i+w_{i+1}}
\label{v-renormaliztion-transform-matome}
\end{equation}
\begin{equation}
{P'}_{i'}(0) = P_i(0)+\frac{w_{i-1}}{w_{i-2}+w_{i-1}}P_{i-1}(0)+
\frac{w_i}{w_i+w_{i+1}}P_{i+1}(0).
\label{initial-condition-renormaliztion-transform-matome}
\end{equation}
\end{itemize}

\subsection{Evaluation of the fixed point}
\label{sec-fixed-points}

At first, we consider the fixed point of the jump rate $w_i$. 
Equation\ (\ref{transition-rate-renormalization-transform-matome}) is 
converted into another form as 
\begin{equation}
\frac{1}{{w'}_{i'}} = \frac{1}{2}\left( \frac{1}{w_i}+\frac{1}{w_{i+1}} 
\right).
\end{equation}
It means that the renormalization transform is equivalent to averaging of the 
inverse of the neighboring jump rates. 
Hence, the global stability of the fixed point is obvious. 
One period of the diffusion constant with $N_l$ sites is renormalized into 
one grid point and the fixed point is reached, 
when the renormalization transform is iterated 
$n_l \equiv \log_2 l/a \equiv \log_2 N_l$ times. 
At the fixed point, the medium is uniform and 
the fixed point value of the jump rate, $w^*$, is independent of the site. 
The value is given by 
\begin{equation}
w^* = 1/\left( \frac{1}{N_l} \sum_{i=0}^{N_l-1} \frac{1}{w_i}\right).
\label{transition-rate-fixed-point}
\end{equation}

Next, we calculate the value of $v_i$ transformed $n_l$ times, $v^*$. 
The list of the renormalization transform of $v_i$ for one period is given 
as 
\begin{eqnarray}
{v'}_{1} & = & v_2+\frac{w_{1}(v_{1}+1)}{w_{N_l}+w_{1}}+
\frac{w_2(v_{3}+1)}{w_2+w_{3}}, \nonumber \\
&& \vdots \nonumber \\
{v'}_{i'} & = & v_i+\frac{w_{i-1}(v_{i-1}+1)}{w_{i-2}+w_{i-1}}+
\frac{w_i(v_{i+1}+1)}{w_i+w_{i+1}}, \nonumber \\
{v'}_{i'+1} & = & v_{i+2}+\frac{w_{i+1}(v_{i+1}+1)}{w_{i}+w_{i+1}}+
\frac{w_{i+2}(v_{i+3}+1)}{w_{i+2}+w_{i+3}}, \nonumber \\
&& \vdots \nonumber \\
{v'}_{N_l/2} & = & v_{N_l}+\frac{w_{N_l-1}(v_{N_l-1}+1)}{w_{N_l-2}+w_{N_l-1}}+
\frac{w_{N_l}(v_{1}+1)}{w_{N_l}+w_{1}}.
\end{eqnarray}
Summing up the both sides respectively, 
the conservation law that the sum of $v_i+1$ over one period does not 
change by the renormalization transform is obtained; 
\begin{equation}
\sum_{i'=1}^{N_l/2} ({v'}_{i'}+1) = \sum_{i=1}^{N_l} (v_i+1).
\end{equation}
Application of the conservation law $n_l$ times gives 
\begin{equation}
v^*+1 = \sum_{i^{(n_l-1)}=1}^{2}(v^{(n_l-1)}_{i^{(n_l-1)}}+1) = 
\cdots = \sum_{i=1}^{N_l} (v_i+1) = \sum_{i=1}^{N_l} 1 = N_l,
\end{equation}
where the fact that the initial value for $v_i$ is given as $v_i=0$ is used.
The fixed point value is given as 
\begin{equation}
v^* = N_l-1.
\label{v-fixed-point}
\end{equation}

There is another parameter $r_i$. After iteration of the transform with $n_l$ 
times, the value of the parameter, $r^*$, is independent of location of 
the grid point. Hence, the jump rate $w_i(s)$ is transformed into 
$w^*(s^*)=w^*+r^* s^*$. We show below that the second term can be neglected. 
Since $s^*=N_l s, s/w^* \sim (a/L_0)^2$ from 
eqs.\ (\ref{order-estimation}, \ref{laplace-variable-rescaling-matome}), 
the order of $r^* s^*$ is estimated as 
\begin{equation}
\frac{r^* s^*}{w^*} \sim r^* N_l \left( \frac{a}{L_0} \right)^2 
\sim r^* \frac{l}{L_0}\frac{a}{L_0}.
\end{equation}
Furthermore, the transform formula of $r_i$ and the fixed point of $v_i$ 
(eqs.\ (\ref{r-renormaliztion-transform-matome}, \ref{v-fixed-point})) 
show that 
\begin{equation}
r^* \sim v^* \sim \frac{l}{a}
\end{equation}
Hence, 
\begin{equation}
\frac{r^* s^*}{w^*} \sim \left( \frac{l}{L_0} \right)^2.
\label{ignoring-r}
\end{equation}
It means that $r^* s^*$ can be neglected compared to $w^*$ when 
the macroscopic limit $l/L_0 \rightarrow 0$ is taken.

We discuss the initial condition renormalized $n_l$ times.
Here, for simplicity, we assume that the initial distribution is localized 
at the origin of the chain; $P_i(0)=\delta_{i,0}$.
As a special case of eq.\ (\ref{renormalized-initial-condition}), 
the fixed point is given as 
\begin{equation}
{P^*}_{i^*}(0) \equiv {P^{(n_l)}}_{i^{(n_l)}}(0) = \delta_{i^*, 0}.
\label{initial-condition-fixed-point}
\end{equation}

From eqs.\ 
(\ref{laplace-variable-rescaling-matome}, \ref{space-index-rescaling-matome}, 
\ref{laplace-transform-probability-rescaling-matome}), 
the values of $i, s, P_i(s)$ rescaled $n_l$ times are given as 
\begin{eqnarray}
i^* & = & i/N_l \nonumber \\
s^* & = & N_l s \nonumber \\
{P^*}_{i^*}(s^*) & = & P_i(s)/N_l 
\label{variables-fixed-point}
\end{eqnarray}

\section{The Effective Diffusion Equation}
\label{sec-eff}

The master equation at the fixed point is written as 
\begin{equation}
(v^*+1) s^* {P^*}_{i^*}(s^*)-{P^*}_{i^*}(0) 
\simeq w^* \left[
{P^*}_{i^*-1}(s^*)-2 {P^*}_{i^*}(s^*)+{P^*}_{i^*+1}(s^*) \right], 
\end{equation}
where the symbol $\simeq$ denotes equality when the limits 
$a/L_0\rightarrow 0, l/L_0\rightarrow 0$ are taken. 
The former limit is the continuum limit and the latter corresponds to the 
large spatial scale and the long time limit $D(x)t/l^2 \rightarrow \infty$.

With the fixed point values of 
eqs.\ (\ref{v-fixed-point}, \ref{initial-condition-fixed-point}, 
\ref{variables-fixed-point}), the master equation is rewritten in terms of 
the non-renormalized parameters as 
\begin{equation}
N_l s P_i(s)-P_i(0) \simeq \frac{w^*}{N_l}
\left[P_{i-N_l}(s)-2 P_i(s)+P_{i+N_l} \right].
\end{equation}
The inverse Laplace transform gives 
\begin{eqnarray}
\frac{dP_i(t)}{dt} & \simeq & \frac{w^*}{N_l^2}\left[ P_{i-N_l}(t)-2P_i(t)
+P_{i+N_l}(t) \right] \nonumber \\
& \simeq & \frac{a^2 w^*}{l^2} \left[ P_{i-N_l}(t)-2P_i(t)+P_{i+N_l}(t) \right]
\end{eqnarray}

The continuum limit is taken by dividing the both sides by $a$ and 
the limit $a\rightarrow 0$ is taken. By using the 
definition $\rho(x,s) \equiv \lim_{a\rightarrow 0} P_i(s)/a$
(eq.\ (\ref{definition-discrete-probability})) in the procedure, 
\begin{equation}
\frac{\partial \rho(x,t)}{\partial t} \simeq \frac{D_e}{l^2}
\left[ \rho(x-l,t)-2 \rho(x,t)+\rho(x+l,t) \right]
\end{equation}
is obtained.
Here, we introduced the new parameter 
\begin{equation}
D_e \equiv \lim_{a\rightarrow 0}a^2 w^*.
\label{definition-effective-diffusion-constant}
\end{equation}
Since spatial scale of observation is of order of $L_0$, the spatial 
coordinate $x$ is changed to $\bar{x}\equiv x/L_0$ as 
\begin{equation}
\frac{\partial\rho(\bar{x},t)}{\partial t} \simeq \frac{D_e}{{L_0}^2} 
\frac{\rho(\bar{x}-l/L_0,t)-2 \rho(\bar{x},t)+\rho(\bar{x}+l/L_0,t)}
{(l/L_0)^2}.
\end{equation}
The limit $l/L_0\rightarrow 0$ is taken with 
the observation scale $L_0$ is fixed at a finite value.
The result of the limit is given as 
\begin{equation}
\frac{\partial \rho(\bar{x},t)}{\partial t} = \frac{D_e}{{L_0}^2} 
\frac{{\partial}^2 \rho(\bar{x},t)}{\partial {\bar{x}}^2},
\end{equation}
where we assume that the distribution of the random walker after long-time 
development is smoothed away and the derivative on the right hand side exists.
Since $L_0$ is finite and the spatial variable can be changed back to 
$x\equiv L_0 \bar{x}$, the effective diffusion equation is obtained as 
\begin{equation}
\frac{\partial \rho(x,t)}{\partial t} = D_e 
\frac{{\partial}^2 \rho(x,t)}{\partial x^2}.
\label{effective-diffusion-equation}
\end{equation}

It is important to note that the solution of the effective diffusion 
equation, $\rho(x,t)$, is the non-averaged density. 
It means that the solution to the original diffusion equation (eq.\ 
(\ref{diffusion-equation})) at time $t \gg l^2/D(x)$ satisfies the effective 
diffusion equation without any data-processing like averaging.

Finally, we evaluate the effective diffusion constant defined by 
eq.\ (\ref{transition-rate-fixed-point}). The continuum limit gives 
\begin{eqnarray}
D_e & \equiv & \lim_{a \rightarrow 0} a^2 w^* \nonumber \\
& = & \lim_{a \rightarrow 0} \frac{l}{\frac{l}{N_l}\sum_{i=0}^{N_l-1} 
\frac{1}{D_i}} \nonumber \\
& = & l/\left[ \int_0^l dx \frac{1}{D(x)} \right].
\label{effective-diffusion-constant}
\end{eqnarray}

The formula, eq.\ (\ref{effective-diffusion-constant}), implies that 
the effective diffusion constant is mainly determined by regions 
where the value of the diffusion constant is small. 
In other words, the speed of the diffusion process is limited by the region 
where diffusion is slow. It can be called the bottle-neck effect.

\section{Discussion and Summary}
\label{sec-summary}

In this paper, the effective diffusion constant of diffusion process in 
periodic inhomogeneous media is evaluated analytically with the 
real-space RG method. 

As pointed out in \S\ \ref{sec-intro}, the problem has been solved with 
the multi-scale method. We compare the two methods. 

In the multi-scale method, different space-time scales 
$x, t; \epsilon x, \epsilon^2 t; \ldots$ characterized by 
arbitrary small number $\epsilon$ 
are introduced and the perturbation expansion in terms of $\epsilon$ is 
performed. 
The parameter $\epsilon$ describes separation of the two characteristic 
scales, the scale of the microscopic structure of the medium and the 
observation scale. 
Hence the parameter $\epsilon$ corresponds to the parameter $L_0$ used to 
characterize the scale of observation in the RG method. 
The development equations of the slow variations are obtained as 
conditions for absence of singular and secular terms, which are called 
the solvability conditions and compatibility conditions. 

In general, the proper slow variables, $\epsilon x, 
\epsilon^2 t; \epsilon^2 x, \epsilon^4 t; \ldots$, cannot be selected 
automatically and selection needs trial and error. On the other hand, 
in the RG methods the slow variables are obtained automatically as 
rescaled variables $x^*, t^*$ at the fixed points \cite{footnote2}. 
However, the extension as explained in \S\ \ref{sec-extension} and the 
decimation procedure have to be performed in a proper way so that the 
renormalization transform is as simple as possible. 
Selection of the best way of extension and 
decimation needs some trial and error. 

Furthermore, the renormalization transform is formulated as recursion 
formulae of parameters. Hence, even if the transform is so complicated that 
evaluating the fixed point analytically is difficult, it would be easier to 
compute the fixed points numerically. 
Especially, the real-space RG on a discrete 
lattice space can be converted to numerical algorithms easily. 

Although the RG method has some advantages, for the problem studied in the 
present paper, calculation with the multi-scale method is much easier than 
that with the RG method.

\section*{Acknowledgments}
We thank Prof.\ Sasa for letting us know the interesting problem and 
valuable discussions. 
We wish to acknowledge valuable discussions with Prof.\ Nakano, 
Prof.\ Kaneda, Prof.\ Gotoh, Prof.\ Okamura, Prof.\ Fujisaka, 
Prof.\ S.-I. Itoh, Prof.\ Yagi, Mr.\ Kitahara, Mr.\ Furuya and Dr.\ Kitazawa. 
This work was partly supported by the Grant-in-Aid for Scientific Research of 
Ministry of Education, Culture, Sports, Science and Technology.


\begin{thebibliography}{99}
\bibitem{footnote-discretization}Although it is more natural that time is also 
discretized and coarse-grained, the variables at different times would appear 
in renormalization transform and hence the renormalization transform would 
become complicated.
\bibitem{footnote2}The advantage is possessed in common by the method of 
ref.\ 6.
\end{thebibliography}
\end{document}